# Towards a Systematic Approach for Smart Grid Hazard Analysis and Experiment Specification

Paul Smith, Ewa Piatkowska, Edmund Widl, Filip Pröstl Andrén, Thomas I. Strasser
AIT Austrian Institute of Technology, Vienna, Austria
{firstname.lastname}@ait.ac.at

*Abstract*—The transition to the smart grid introduces complexity to the design and operation of electric power systems. This complexity has the potential to result in safety-related losses that are caused, for example, by unforeseen interactions between systems and cyber-attacks. Consequently, it is important to identify potential losses and their root causes, ideally during system design. This is non-trivial and requires a systematic approach. Furthermore, due to complexity, it may not be possible to reason about the circumstances that could lead to a loss; in this case, experiments are required. In this work, we present how two complementary deductive approaches can be usefully integrated to address these concerns: Systems Theoretic Process Analysis (STPA) is a systems approach to identifying safety-related hazard scenarios; and the ERIGrid Holistic Test Description (HTD) provides a structured approach to refine and document experiments. The intention of combining these approaches is to enable a systematic approach to hazard analysis whose findings can be experimentally tested. We demonstrate the use of this approach with a reactive power voltage control case study for a low voltage distribution network.

*Index Terms*—Experiment Specification, Hazard Analysis, Holistic Test Description, Smart Grid, STPA-SafeSec.

## I. INTRODUCTION

Electric power systems are being transitioned to the smart grid [1]. This involves the extensive use of Information and Communication Technology (ICT) to support new services, including new forms of local and distributed control as well as Renewable Energy Resources (RES) [2], [3]. Building on these new ICT systems, these control strategies can use distributed sensor data to inform the choice of control actions that are applied to similarly distributed power systems. Moreover, the timescales that control behaviour is realized varies, from short-term local control through to longer-term strategies, such as demand-response schemes. The result is a relatively complex arrangement of local and increasingly distributed control strategies that aim to ensure the safe and secure operation of the grid in the presence of dynamic generation and loads.

The transition to the smart grid has many benefits. However, the increased complexity it introduces and the use of ICT has the potential to result in safety-related losses, such as blackouts; this is for various reasons. These reasons may be benign – for example, unexpected interactions between controllers or changing grid conditions – from those anticipated at design time – could result in a loss. Meanwhile, the use of ICT systems introduces the increased risk of losses that have malicious origins, i.e. cyber-attacks. There is academic literature and recent real-world examples that have shown how cyber-attacks could result in grid-related losses [4].

With this in mind, it is beneficial to identify potential losses and their causes while the system – or a new control strategy that is being applied to an existing system – is being designed and engineered. This is the basic tenet of advocates of performing *security by design*, for example. However, because of the complexity of the systems concerned, this activity is non-trivial. For example, it may not be readily deducible how a set of local and distributed control strategies will interact – potentially leading to a loss – or be adversely affected by an adversary. Consequently, a systematic approach is required to identify the losses of concern and the scenarios that could yield them. Moreover, it may be necessary to perform experiments to understand the precise nature of the scenarios that result in a loss. With the results from these experiments, design and engineering decisions can be made to mitigate any flaws (safety) or vulnerabilities (cybersecurity) that have been identified.

To this end, we propose to integrate two complementary approaches that help to address the dual concerns of *(i)* performing a hazard analysis to identify losses and their causes; and *(ii)* specifying experiments that can be used to test the non-trivial findings from a hazard analysis. For the former, we advocate the use of the Systems Theoretic Process Analysis (STPA) approach from Leveson [5]. This is a top-down systems approach to identifying losses and their causes, resulting in the specification of so-called hazard scenarios. Meanwhile, for the latter, we propose the use of the ERIGrid power system testing approach and corresponding Holistic Test Description (HTD) [6], which provides a structured approach to refining and documenting experiments. The complementarity of these approaches stems from them both applying a deductive top-down approach, with the results from an increasingly specific STPA analysis providing input for similarly increasingly specific test and experiment specifications.

To indicate the usefulness of our combined approach, we present a voltage control case study for a low voltage power distribution network. For this study, we use an extension of STPA, called STPA-SafeSec [7], which accounts for implementation-oriented cybersecurity concerns. STPA-SafeSec can be used to support the specification of experiments that examine the potential consequences of cyber-attacks to a target system. The case study shows how the top-down STPA-SafeSec analysis can be used to inform the specification of an HTD, resulting in detailed test and experiment specifications. The intention is that the findings from experiments can be used to inform improved, i.e. more safe and secure, system designs.

This paper is organized as follows: Section II provides an overview of related work, for example, on hazard analysis techniques. Subsequently, the two approaches that are integrated are summarized and overview of how they can be applied in combination is discussed in Section III. We present a case study in Section IV to illustrate the approach and indicate its usefulness. Finally, conclusions and an outlook about potential future activities are summarized in Section V.

## II. RELATED WORK

There are several approaches to performing a hazard analysis, including the Failure Modes and Effects Analysis (FMEA) process [8], a HAZard and OPerability (HAZOP) study [9], and the STPA method [5]. In various ways, these approaches can be used to identify hazard scenarios based on a high-level design of a system. However, they typically do not consider the underlying technical implementation – i.e. computer systems and networks – as part of the analysis. For the purpose of specifying lab-based experiments, a systematic approach to relating high-level losses and hazards – caused by failures in control – to technical systems that could be exploited to yield a controller failure is important. To this end, we have chosen the STPA-SafeSec method [7], which supports this mapping.

At the same time, due to the further integration of ICT solutions and other energy systems into smart grid applications, there is an increased need for new test and system validation processes that goes beyond the scope of cybersecurity [10].

Device-level testing approaches are common and wide-spread in the power systems domain [11], but when it comes to an integrated view on the system level – which is the case for smart grid applications – only a few approaches exist [12]. One of such few approaches is the JRC Interoperability Testing Methodology, which focuses mainly on the evaluation of ICT interoperability of connected devices and services [13]. Another method applies the well-know Hardware-in-the-Loop (HIL) technique to whole power system configurations [14] but it focuses on the test execution and does not provide a systematic approach for analysing and specifying test cases and experiment specifications. To this end, a methodology for framing a holistic approach to testing has been developed in the ERIGrid project [15], in order to capture this complexity. This methodology, referred to as the *Holistic Test Description* (HTD) [6], aims at enabling the testing of new solutions within their relevant operational context. This testing methodology ensures a clear vocabulary for smart grid testing across engineering disciplines and a common understanding of how to describe a testing approach that addresses a system-relevant perspective. However, HTD is a general approach for describing test cases whereas the analysis of hazardous scenarios is not in the focus.

Summarizing, none of the above outlined approaches have an integrated, multi-domain view of smart grid systems taking hazard analysis and experiment design into account.

## III. PROPOSED INTEGRATED APPROACH

In this section, we briefly introduce the two approaches that are integrated – STPA-SafeSec and the ERIGrid Holistic Test Description. Subsequently, we provide a summary of how the two approaches can be integrated.

### A. STPA-SafeSec Overview

STPA-SafeSec is an extension of the STPA hazard analysis approach. It extends STPA in several ways to support cyber-security aspects. A high-level overview of the steps in the STPA-SafeSec process are presented in Fig. 1 – several of these steps are adopted from standard STPA.

Initially, *losses* and *hazards* are defined (not shown in Fig. 1). Losses describe events that should not occur, such as interruption of supply or voltage violations, whereas hazards are high-level scenarios that could result in losses. Moreover, *constraints* that should be not violated are defined – this can be achieved by negating the specified hazards, i.e., a constraint is the system should *not* be in a defined hazardous state. Subsequently, for each identified control loop in the system, a high-level definition of the control layer is specified, using a generic control structure diagram. Using this diagram and the specified hazards, *hazardous control actions* are identified – these describe a control action, a system state, a definition of when (or not) the application of a control action could result in a hazard given a system state, and the resulting hazard (see Table II for an illustrative example).

STPA-SafeSec then requires the mapping of the control layer onto the underlying components that realize them. This is achieved with the support of a *generic component layer diagram*, in a similar manner to the approach at the control layer. With this mapping, generic threats can be enumerated that may result in safety constraints being violated. Finally, hazard scenarios are defined, which collate the information that has been identified in the previous steps.

### B. ERIGrid Holistic Test Description

The HTD approach aims to support domain experts in writing up intentions and drawing out configurations, with the goal to identify and define the essential parameters and procedural steps for conducting a test. It comprises a set of textual templates [16], a graphical notation and partial processes that may be employed by a practitioner to structure, refine and document their testing endeavour [6], [16].

In HTD, the *System under Test* (SuT) identifies the abstract, categorical, system boundaries of an abstract test system encompassing all relevant sub-systems and interactions (domains) required for the investigation. The *Object(s) under Investigation* (OuI) identify the subsystem(s) or component(s) in scope of the test objective, and with respect to which the test criteria need to be formalized. The *Domain(s) under Investigation* (DuI) identify the relevant physical or cyber-domains of test parameters and connectivity, which can be listed directly or formulated in the form of a hierarchy. With reference to *Use Cases*, the full set of *Function(s) under Test* (FuT), and the specific *Function under Investigation* (FuI) are identified.

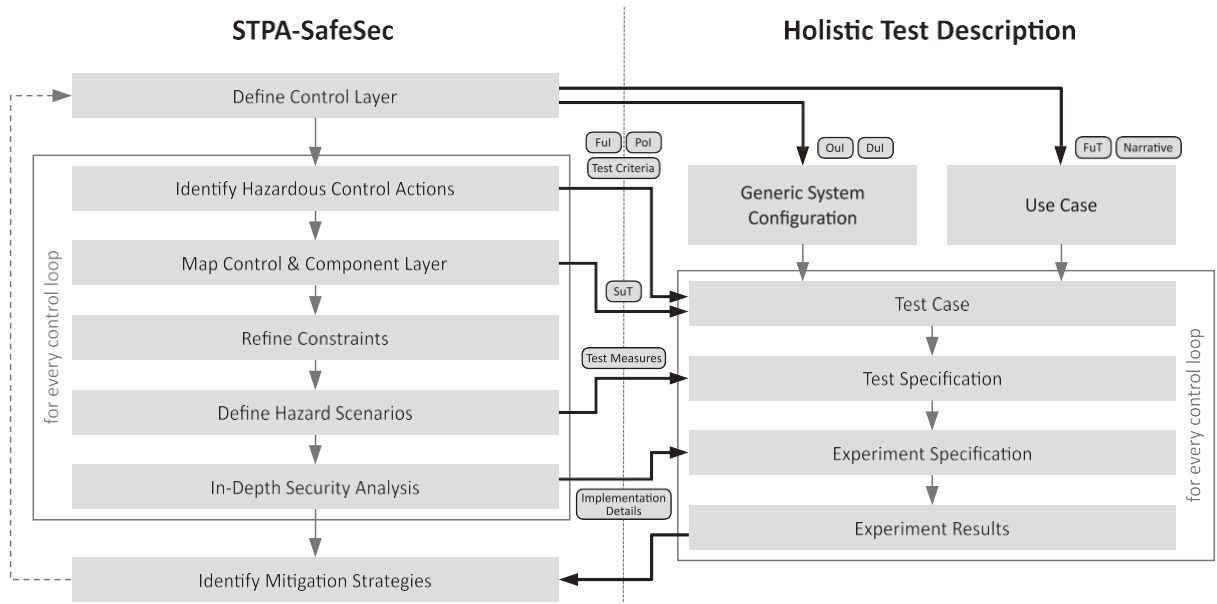

Fig. 1: Mapping between STPA-SafeSec and the Holistic Test Description

The *Purpose of Investigation* (PoI) formulates the test objective, also stating whether it relates to characterization, validation or verification objectives. Together the above items inform the *Test Criteria*, which formalize the test metrics into target criteria, variability attributes, and quality attributes (which are typically thresholds for acceptable results).

The HTD introduces three main levels of test definitions, where each references the previous level, leading to an incremental scoping of a concrete test/experiment:

1) A *Test Case* (TC) provides a set of conditions under which a test can determine whether or how well a system, component or one of its aspects is working given its expected function.
2) A *Test Specification* (TS) defines the test system (i.e. how the OuI is to be embedded in a specific SuT), which parameters of the system will be varied and observed for the evaluation of the test objective, and in what manner the test is to be carried out (test design).
3) The *Experiment Specification* (ES) defines by what exact means a given TS is to be realized in a given laboratory infrastructure or simulation implementation.

A TC formulates key objectives and context of a test, whereas the TS and the ES provide a concrete foundation for the actual test execution.

### C. Integrating STPA-SafeSec with ERIGrid HTD

The approaches of STPA-SafeSec and ERIGrid HTD can be combined to support joint workflows involving domain experts for safety, cybersecurity and smart grid engineering. On the one hand, STPA-SafeSec provides the tools to identify potentially hazardous scenarios, which could include cyber-attacks. On the other hand, the HTD provides a structured approach to break down and translate these findings into concrete lab experiments.

Fig. 1 shows how both approaches complement each other. When applying STPA-SafeSec, all relevant information for specifying TCs and TSs can be retrieved, which can be subsequently used to design and and conduct relevant lab experiments (via the derived ES). In fact, the overlaps between both methodologies are such that the outcomes of STPA-SafeSec can be gradually mapped to the HTD, going from the more generic outlines (e.g. the FuT or OuI) to application-specific high-detail test designs (e.g. test criteria, SuT and target measures). As such, the HTD supports the STPA SafeSec approach by easing the design of supplemental lab experiments, which can provide insights about system-specific artefacts that are hard or sometimes even impossible to trace from theoretic considerations alone.

## IV. VOLTAGE CONTROL CASE STUDY

To illustrate the application of the integrated approach, we present a smart grid voltage control case study, which was developed in the ERA-Net-funded LarGo! project[1]. This case study is depicted in Fig. 2 and concerns voltage control in a smart low-voltage distribution network. As can be seen in Fig. 2, there are three control loops that are being used to manage voltage levels: *(i)* control within the substation, wherein a local controller adjusts voltage levels using an On-Load Tap Changer (OLTC); *(ii)* signalling from the substation to distributed Building Energy Management Systems (BEMS) to curtail power injection to the grid; and *(iii)* reactive power control, in which photovoltaic (PV) inverters are being used to manage reactive power levels at distributed locations. A hazard analysis of this scenario has been performed, as part of the LarGo! project [17]; here, we provide highlights from

---
[1]http://www.largo-project.eu/

TABLE I: High-level losses and hazards for the voltage control case study.

(a) High Level Losses

| | Losses |
|---|---|
| L1 | Damage to power equipment |
| L2 | Interruption of power supply to consumer loads |
| L3 | Loss of service |
| L4 | Financial loss |
| L5 | Grid instability |

(b) Hazards to Losses Mapping

| | Hazards | Related Accidents |
|---|---|---|
| H1 | Inability to allocate in-feed power (over voltage) | L1,L5 |
| H2 | Inability to meet local demand (under voltage) | L2,L3 |
| H3 | Voltage Oscillations | L1,L5 |
| H4 | Suboptimal service operation | L4 |
| H5 | Inability to control | L1,L2,L3 |

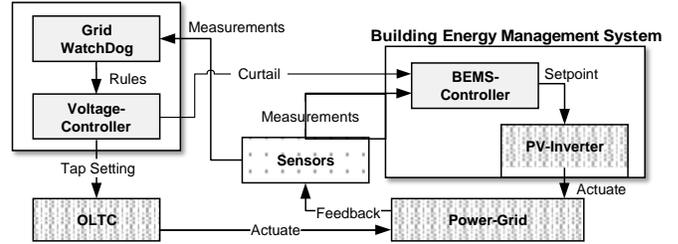

| Control | Control signal hazardous when applied: | | | | Label |
|---|---|---|---|---|---|
| | Any time | Too early | Too late | Not | |
| Max injection | (H1) | | | | HC-1 |
| Injection | (H1) | | H3 | | HC-2 |
| Neutral | | | H3 | | HC-3 |
| Consume | | | H3 | (H1) | HC-4 |
| Max consume | | | | (H1) | HC-5 |

Consequently, it may be desirable to define tests and perform experiments to explore these parameters.

*C. Identify Causal Factors*

The next step is to determine the *causal factors* that could result in the hazardous control actions. These causal factors capture high-level deficiencies that could result in a hazardous control action being applied. To support this activity, Leveson defines a generic causal factor diagram, which can be adapted to the control narrative being analysed (see [5], p. 223). An extract of the causal factor diagram that has been tailored to the PV-Inverter control scenario is presented in Fig. 3.

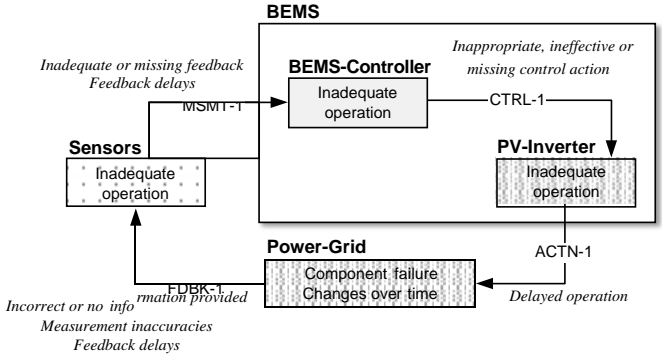

Fig. 3: An extract of the factors that could cause hazards for local PV-Inverter control [17].

As an example, consider the causal factors that could result in hazardous control actions *HC-2* through to *HC-4* being applied; specifically, those that can result in voltage oscillations (*H3*), as a consequence of a setpoint being applied too late. Examining Fig. 3, it can be deduced that the causal factors for these actions relate to potential issues with the feedback loop from the *Power-Grid* to the *BEMS-Controller*, caused by 'feedback delays' at *FDBK-1* and *MSMT-1*, and 'inadequate operation' of the *Sensors* and *BEMS-Controller*. In practice, these causal factors could be the result of poor implementation choices, e.g. using under-provisioned communication network technology to send measurements to the *BEMS-Controller*.

STPA-SafeSec introduces a further step here, wherein the control layer is mapped onto its underlying component implementation. The purpose of this step is to enable *security constraints* [7] to be identified that relate to its implementation. Violation of these security constraints can result in the identified causal factors, such as feedback delays. This step is supported, in a similar way to the casual factor analysis, with a generic component layer diagram and a set of generic cybersecurity threats that can be assigned to components, such as devices and networks. An implementation of a home energy management system has been proposed by Shakeri *et al.* [19], which relies on wireless communication to interconnect devices in a Home-Area Network (HAN). In an analysis, the causal factor diagram (Fig. 3) would be mapped onto a corresponding component diagram, not shown here, wherein *MSMT-1* is realised by a wireless network. Using the generic cybersecurity threats that have been defined by Friedberg *et al.*, this communication could be subject to a *communication delay* threat (*CSTR-A-1*) – see Table 2 in [7]. In practice, this threat could be realized via wireless jamming, resulting in a Denial of Service (DoS) attack [20], for example.

*D. Define Hazard Scenarios and HTD Test Specifications*

Next *hazard scenarios* are defined, which collate the information that has been gathered in the previous steps. They describe the causal factors that can result in a hazardous control action being applied, potentially leading to a hazard and a loss. An example hazard scenario is presented in Table III. It collates the information that has been gathered for loss *L5* and hazard *H3* that are associated with voltage oscillations, caused by delayed application of the reactive power setpoint.

TABLE III: An example hazard scenario for the reactive power case study.

---

**Hazard Scenario:** *HS-1: Reactive Power Control is Inhibited Due to Feedback (Measurement) Delay*
**Losses and Hazards:** H3 resulting in L5 (see Tab. I)
**Hazardous Control Actions:** HC-2 to 4 (see Tab. II)
**Control Level Entities:** FDBK-1 or MSMT-1 (see Fig. 3)
**Causal Factors:** Feedback delay at MSMT-1 or FDBK-1
**Component Level Entities:** HAN connection from *Sensors* to *BEMS*
**Security Constraints:** CSTR-A-1 (Communication delay) [7]
**Safety Constraints:** The system should not create voltage oscillations

---

In our integrated approach, hazard scenarios are used to support the definition of *Test Specifications*. An example specification is shown in Table IV. The mapping between hazard scenarios and test specifications (i.e. is there a 1:1 or 1:N mapping) can be arbitrarily defined, as convenient. In the example shown, there is a 1:1 mapping between the scenario and the test description.

A test description includes a *title* and a *rationale*. In this case, the rationale for the test is to examine the severity of losses that could result from hazard scenario *HS-1*. This involves exploring the parameters of the scenario, including the affect of the duration of the feedback delay, and the number and location of BEMSs. The *test system* is specified – at the level of abstraction that is required for a test specification, the information that has been used for the STPA analysis is sufficient. (Further technical details are required for the experiment specification.) For the specified test, the *target measures* that will be examined are the voltages and their characteristics, both local to a targeted (set of) BEMS and the wider distribution line.

Subsequently, *input and output parameters* are defined – these specify the controllable parameters that will be *changed* (in experiments), those that will remain *unchanged*, and the *measured* parameters. In the example test specification, it is defined that there will be tests that are used to measure the affect of different feedback delays $d$, the target BEMS $v_i$ where the delay is introduced (where $i \leq N$, and $N$ is the number of BEMSs on the line) and the number of BEMSs in a test $n$. The configuration of load profiles and the droop law that

TABLE IV: An example HTD Test Specification derived from the hazard scenario HS-1, presented in Table III.

| Test Title | Assessment of the loss *L5* due to reactive power control being inhibited, caused by feedback (measurement) delay *(HS-1)* |
|---|---|
| Rationale | Examine the potential severity and parameters that are related to HS-1 (see Tab. III.) |
| Test System | See Fig. 2 and description in Sec. IV |
| Target Measures | Voltages and their dynamics, i.e. oscillations, both local to a target BEMs and the wider distribution line. |
| I/O parameters | *Controllable Input Parameters*<br>- Feedback delay $d$ (e.g. $d = 100ms, 1s$)<br>- Target ($v_i$) and number ($n$) of BEMS<br>*Uncontrollable Parameters*<br>- Power consumption of loads<br>- Droop law configuration<br>*Measured Parameters*<br>- Voltage at loads |
| Test Design | - Controllers at the BEMS are configured with an appropriate droop law that dictates their reactive power behaviour<br>- A test begins with the grid operating under normal conditions (see *Initial System State*), with the BEMS configured to provide power to the grid.<br>- For each test, a measurement delay $d$ is introduced to the voltage measurements that are provided to a target (set of) *BEMS-Controller* (see Fig. 3)<br>- Depending on the hazardous control to be tested, i.e. HC-2 to 4, the voltage level at the target BEMS should be such that the reactive power set-point of the inverter is *different* from the targeted hazardous control action.<br>- To test whether the hazardous control action results in a loss (i.e. L5), the voltage level at the BEMS should be changed, e.g. by increasing the local load, such that the targeted hazardous control action should be applied by the controller (*BEMS-Controller*).<br>- Voltage levels are measured at all loads to determine the characteristics of the potential oscillations that are caused by introducing feedback delay<br>- Variations on the basic test can include changing the target and number of BEMS, to determine whether location on the line affects the magnitude of loss |
| Initial System State | Nominal grid state (i.e. $v_g = 230V \pm 10\%$) with BEMS providing power to the grid; $d = 0$ |
| Source of uncertainty | Feedback delay ($d$), the target BEMS ($v_i$) and ($n$) number of BEMS |

is used by the *BEMS-Controller* are not controlled, i.e. not changed, during tests. The measured parameter in tests is the voltage at loads, to determine the characteristics of potential voltage oscillations (*H3*) and grid instabilities (*L5*)

The *test design* describes the steps that will be taken to perform a test. Initially, the grid is configured appropriately (e.g. using the uncontrolled parameters) and is operating under normal conditions. In this test, normal conditions are defined by the *initial system state*: the grid-wide voltage $v_g = 230V \pm 10\%$ and BEMSs are providing power to the grid. Furthermore, the feedback delay $d = 0$. The hazard scenario *HS-1* relates to three hazardous control actions (*HC-2* through to *HC-4*), in which a different reactive power setpoint is applied too late. To test the potential effect of applying a setpoint too late, the *BEMS-Controller* must respond to a change in the measured voltage that requires a new setpoint to be applied. To achieve this, the measured voltage must change sufficiently to cause a new setpoint to be applied. This can be achieved, for example, by changes in the load profile of the target BEMS $v_i$, either by defining a dynamic load profile or introducing a change that causes the target setpoint to be applied. The delay $d$ is applied to the voltage measurement that is supplied to the *BEMS-Controller*, such as $d = 100ms, 1s$. As mentioned earlier, voltage is measured throughout the grid and variations of the basic test can be applied by changing the target and number of BEMS.

### E. In-Depth Security Analysis and Experiment Specification

The penultimate step in STPA-SafeSec is to perform a in-depth security analysis, which involves identifying the cybersecurity vulnerabilities and the threats that are associated with the system implementation. This security analysis can be used to inform the detailed specification of experiments using the HTD approach. For example, for the test specification outlined in Table IV, different experiments could be defined that realize, e.g. via simulation, the generic availability threat (*CSTR-A-1*), such as a wireless jamming attack [20], a DoS attack that targets the *Sensor*, or a so-called JellyFish attack [21], if a multi-hop network is used to connect devices in the HAN. The intention is to characterize how vulnerable the system is to the identified threats, and whether they could lead to the concerned losses.

Finally, experiments are performed that use the detailed experiment specifications. The findings from these experiments can be used to inform suitable mitigation strategies that address the causal factors, which have been identified in the STPA-SafeSec analysis. In the case study, it could be appropriate to not use wireless networking technology or introduce systems to detect cyber-attacks that introduce feedback delay.

## V. CONCLUSIONS

In this work, we have presented a novel integration of two complementary approaches – STPA-SafeSec and the ERIGrid HTD – that can be used to analyse the potential losses and hazards that are associated with a smart grid, and populate an increasingly detailed test and experiment specification. The aim is to provide a systematic approach to identify hazard scenarios and facilitate the specification of experiments, which can be used to test the nature of losses that are non-trivial. We have illustrated the usefulness of the approach with a voltage control case study for a smart low distribution network. In this case study, hazard scenarios are identified for a distributed reactive power voltage control strategy. The nature and severity of losses in this study are not readily deducible, as they depend on several factors: the magnitude of feedback delay introduced to measures, the state of other control strategies, e.g. local to a substation and at other loads, and the power grid configuration and state. Consequently, well-designed experiments must be performed to understand the nature of the risk that is associated with a hazard scenario. Our approach facilitates the specification of such experiments, the findings from which can be used to improve system design.

Potential future work should focus on the refinement of the proposed integration of the two selected methods, their application on further case studies, as well as the development of corresponding software tools for easier analysis of smart grid applications.

ACKNOWLEDGMENTS

This work received funding in the European Community's Horizon 2020 Programme (H2020/2014–2020) under project "ERIGrid" (Grant Agreement No. 654113) as well as in the framework of the joint programming initiative ERA-Net Smart Grids Plus under project "LarGo!" (FFG No. 857570).